\documentclass[12pt]{article}

\usepackage{amsfonts,amssymb,amsmath}
\usepackage{graphicx}
\DeclareGraphicsExtensions{.pdf,.png,.jpg}
\begin{document}
\title{\bf Quantum phase estimations with spin coherent states superposition}
\author{ Yusef Maleki $^{1}$
\thanks{E-mail: maleki@physics.tamu.edu}
\\ $^1${\small Department of Physics and Astronomy, Texas A\&M University, College Station, Texas 77843-4242, USA}
}
 \maketitle

\begin{abstract}
The quantum metrological performance of spin coherent states superposition is considered, and conditions for measurements with the  Heisenberg-limit (HL) precision are identified. It is demonstrated that the choice of the parameter-generating operator can lead to physically different estimation outcomes. In particular,  closed-form analytical descriptions for the performance of spin cat states are derived.
These findings show the routes to careful control of parameters necessary for achieving HL precision and provide insightful information on the geometry of the specific coherent state superposition and its relevance to the performance of the states for parameter estimations.
\end{abstract}


\newpage
\section{Introduction}

High-precision measurements are of utmost importance in fundamental physics, and various areas of technology \cite{Giovannetti,Bevington,Holland,Bollinger,Pirandola,Abbott,Ono,Cappellaro}
. In the classical realm, the precision of measurements can be enhanced by repeated measurements. More specifically, with $N$ independent measurements performed to estimate a parameter $\xi$, the error of $\xi$ estimation is bounded by the shot-noise limit, $\Delta\xi\propto\frac{1}{\sqrt{N}}$ \cite{Giovannetti,Bevington}. In contrast, quantum physics provides tools for circumventing this limit \cite{Cappellaro,MalekiRecovery}, enabling measurements with precisions within the limit of  $\Delta\varphi\propto\frac{1}{N}$, referred to as the Heisenberg Limit (HL) \cite{MalekiRecovery,Xiang}. 
\par
As an important example of such improvements, cat states of the form $\frac{1}{\sqrt{2}} (|N\rangle_a
\otimes|0\rangle_b +|0\rangle_a \otimes|N\rangle_b)$, known as N00N states \cite{Boto2000,Dowling2008}, have been shown \cite{Giovannetti,Dowling2008} to provide the HL sensitivity. However, the problem is that such states are  difficult to generate and are extremely prone to decoherence \cite{Xiang,Dowling2008,Afek}.  To resolve this, specific schemes have been proposed \cite{Maleki2018,Maleki2019}.
Alternatively, other quantum states have been considered, which can provide better than shot-noise limit precision. An important example of such a cat state is the superpositions of two SU(2) coherent states, which attracted much attention \cite{Sanders,Huang1,Huang2,Maleki2020}.

Spin cat states are promising candidates for quantum metrology due to their strong capability for beating the shot-noise limit and their robustness against decoherence \cite{Sanders,Huang1,Huang2}.  As an important feature, these cat states can be prepared in a wide variety of settings such as dynamical evolution of $N$ two-level systems \cite{Agarwal1}, Bose-Einstein condensates of two-level atoms \cite{Lucke,Strobel} or two-mode photon system in cavity QED \cite{Maleki2020}. Therefore, parameter estimation with spin cat state is compatible with both Ramsey \cite{Sommer} and  Mach-Zehnder interferometry \cite{Sturm}. In all of these scenarios, the coherent state basis could be considered as a two-mode bosonic-harmonic-oscillator basis  \cite{Sanders,Huang1,Huang2,Maleki2020}.
\par
Here, we consider the quantum metrological performance of spin coherent states superposition and identify conditions for measurements with a  Heisenberg-limit (HL) precision. We demonstrate that the choice of the parameter-generating operator plays a critical role in the performance of such a superposition state and consider physically relevant parameters induced by different generators. These findings suggest that careful control of parameters is necessary for achieving HL precision and provide insightful information on the geometry of the specific coherent state superposition on the Bloch sphere and its relevance to the performance of the states for parameter estimations.





\section{Spin coherent states for phase estimation}

This section briefly reviews the spin (SU(2)) coherent states and their essential properties.
Spin coherent states are defined through the action of  rotation operator \cite{Agrawal}

\begin{equation}\label{rotation}
R(\theta,\varphi) =\exp{\lbrace\frac{-\theta}{2}(J_+
e^{-i\varphi}-J_- e^{i\varphi})\rbrace},
\end{equation}
to the reference state $\vert j,-j\rangle$. The rotation operator in  Eq. (\ref{rotation}) can be identified in terms of the generators of the su(2) algebra. The generators of su(2) algebra are denoted by  $J_-$, $J_+$ and $J_z$.  $J_-$ and $J_+$  are the spin-state lowering and raising operators, respectively, and $J_z$ is the $z$-component of the spin operators. As generators of the su(2) Lie algebra, the operators
$J_\pm$ and $J_z$ satisfy the commutation relations $
[J_+,J_-]=2J_z,[J_z,J_\pm]=\pm J_\pm.$

 The operator in Eq. (\ref{rotation}), acting on the ground state of the Dicke basis, results in  the  spin coherent state \cite{Agrawal,Maleki2}
\begin{align}\label{spin}
|\theta,\varphi, j\rangle=\exp{[\frac{\theta}{2}(J_+
e^{-i\varphi}-J_- e^{i\varphi})]} |j,-j\rangle
=\frac{1}{(1+|\gamma|^2)^j}\sum_{m=-j}^j
\left(
  \begin{matrix}
    2j  \\
    j+m 
  \end{matrix}
  \right)^\frac{1}{2}\gamma^{j+m}|j,m\rangle,
\end{align}
where $\gamma=e^{-i\varphi}\tan(\frac{\theta}{2})$. 

The overlap of two spin
coherent states is given by   \cite{Maleki2}
\begin{align}
\langle \theta_1,\varphi_1, j|\theta_2,\varphi_2, j\rangle=\frac{(1+\bar{\delta}
\gamma)^{2j}}{(1+|\delta|^2)^j(1+|\gamma|^2)^j},
\end{align}
where $\gamma=e^{-i\varphi_1}\tan(\frac{\theta_1}{2})$ and $\delta=e^{-i\varphi_2}\tan(\frac{\theta_2}{2})$, and $\bar{\delta}$ represents the complex conjugate of $\delta$. 
 The connection between the Dicke basis and two-mode bosonic-harmonic-oscillator basis can be understood by considering the  Schwinger realization of the su(2) algebra represented in the two-mode Hilbert space states as
\begin{align}
J_+=a^\dag b,\qquad  J_- =b^\dag a,\qquad  J_z =\frac{1}{2}(a^\dag a -
b^\dag b).
\end{align}
where $a$ and $b$ are the bosonic-harmonic-oscillator annihilation operators of the first and the second modes. While the realization is compatible with various settings, we refer to these operators as the photon annihilation operators.
If the two modes $a$ and $b$ contain $N$-photons in total, taking $j=\frac{N}{2}$ and $m=\frac{N_a-N_b}{2}$, we can express Dicke basis as
$|N_a\rangle |N_b\rangle\equiv |j,m\rangle$ ($N=n_a+n_b$). Applying the $J_\pm$ and $J_z$ operators to the Dicke basis yield $
J_\pm|j,m\rangle=\sqrt{(j\pm m)(j\pm m+1)} |j,m\pm1\rangle$ and
 $J_z|j,m\rangle=m|j,m\rangle.$
Thus, when all of the photons are in one of the two modes, we attain  the two special cases $|0\rangle \otimes|N\rangle\equiv
|j,-j\rangle$ and $|N\rangle \otimes|0\rangle\equiv
|j,j\rangle.$ To demonstrate  the connection of the spin coherent state superposition and metrology we note that the superposition of these two special cases provides the N00N state ($|\textrm{N00N} \rangle= \frac{1}{\sqrt{2}} (|N,0\rangle +|0,N\rangle)$) in the Dicke basis as \cite{Sanders}

\begin{equation}\label{Dicke}
|\textrm{N00N} \rangle= \frac{1}{\sqrt{2}} (|j,j\rangle +|j,-j\rangle).
\end{equation}

Therefore, the N00N state can be considered the superposition of the north and south poles of the Bloch sphere. This particular example demonstrates the utility of the superposition of spin coherent states for quantum metrological purposes. To illustrate this capability, the superposition of coherent states of the form 
\begin{equation}\label{Huang}
|\varphi, j\rangle=\mathcal{N}(|\theta,\varphi, j\rangle+|\pi-\theta,\varphi, j\rangle),
\end{equation}
with $\mathcal{N}$ being the normalization factor, is shown to suppress the shot-noise limit, thus, being of great importance for quantum metrology \cite{Huang1,Huang2}.
However, a more generalized superposition of such spin coherent states can be expressed as
\begin{equation}
|\psi, j\rangle=\mathcal{N}(|\theta_1,\varphi_1, j\rangle+|\theta_2,\varphi_2, j\rangle).
\end{equation}
A natural question is whether these are states beyond the scope of Eq. (\ref{Huang}) that are metrologically useful? If so, what are these states? We are going to consider these questions, in detail, in this work.

Note that the phase shift in N00N state could be achieved by $\exp(i\xi(a^\dag a-b^\dag b)/2)$. This indicates that the path $a$ has gained the phase shift $\xi/2$, while the path $b$ has gained the phase shift $-\xi/2$, resulting in a total phase shift $\xi$. Following this discipline, the phase shift in spin states is usually introduced by $e^{i\xi J_z}$, which is expected from Schwinger realization of the spin algebra.

Given a spin cat state, we would like to consider the metrological performance of the state from a somewhat different perspective. In fact, we ask a natural question of how the spin cat state would perform by introducing parameters through some other operators such as $e^{i\xi J_x}$ or $e^{i\xi J_y}$? 
This may seem to be a pure mathematical curiosity at first sight; however, the question is insightful once we consider it more carefully. In fact, considering the Schwinger representation of the spin operators, $e^{i\xi J_z}$ could be understood as a phase shift in an interferometer, while $e^{i\xi J_x}$  and $e^{i\xi J_y}$ could be considered as beam splitting operators, where the beam splitter transmittance is determined through $\xi$. Therefore, parameter estimation with the latter determines the transmittance of a beam splitter rather than the accuracy of phase shift in an interferometer.





\section{Quantum Metrology with Spin-1/2  Cat States}

We first consider the metrology of spin-1/2 cat states and provide a detailed investigation on the metrological power of such cat states. It is notable that if $j=\frac{1}{2}$, then $\textrm{su}(2)$ algebra
will be generated by Pauli matrices $\sigma_\pm=\sigma_x \pm
i\sigma_y$, and $\sigma_z$. In the Fock state basis, $j=\frac{1}{2}$ corresponds to having only one photon in the coherent state. The coherent state, in this case, could be given as

\begin{align}
|\theta,\varphi, 1/2\rangle=\exp[\frac{\theta}{2}(\sigma_+
e^{-i\varphi}-\sigma_- e^{i\varphi})]
|\frac{1}{2},\frac{-1}{2}\rangle.
\end{align}
Therefore, the state  can be obtained to be 
\begin{align}
|\theta,\varphi, 1/2\rangle=\cos\frac{\theta}{2}|\frac{1}{2},\frac{-1}{2}\rangle+e^{-i\varphi}\sin\frac{\theta}{2}|\frac{1}{2},\frac{1}{2}\rangle.
\end{align}

In this framework, the general form of the superposition of two spin coherent states could be represented as
\begin{equation}
|\text{Cat}, 1/2\rangle=\mathcal{N}(|\theta_1,\varphi_1, 1/2\rangle+|\theta_2,\varphi_2, 1/2\rangle).
\end{equation}
Here, $\mathcal{N}$ is the normalization factor of the spin cat states which can be found to be 
\begin{equation}
\mathcal{N}=\dfrac{1}{\sqrt{2[1+\cos (\frac{\theta_1}{2})\cos (\frac{\theta_2}{2})+\cos (\varphi_1-\varphi_2)\sin (\frac{\theta_1}{2})\sin (\frac{\theta_2}{2})]}}.
\end{equation}
In the Dicke basis, the cat state above can be expressed as
\begin{equation}\label{cat1/2}
|\text{Cat}, 1/2\rangle=\mathcal{N}[(\cos\frac{\theta_1}{2}+\cos\frac{\theta_2}{2} )|\frac{1}{2},\frac{-1}{2}\rangle+(e^{-i\varphi_1}\sin\frac{\theta_1}{2}+e^{-i\varphi_2}\sin\frac{\theta_2}{2})|\frac{1}{2},\frac{1}{2}\rangle].
\end{equation}
We first analyze the performance of the state $|\text{Cat}, 1/2\rangle$ for quantum metrology when the system accumulates a phase through $e^{i\xi \sigma_z}$. 
Once this phase is introduced, the cat state degenerates to 
\begin{equation}\label{cat-1/2}
|\text{Cat}, 1/2\rangle_\xi=\mathcal{N}[e^{-i\xi/2}(\cos\frac{\theta_1}{2}+\cos\frac{\theta_2}{2} )|\frac{1}{2},\frac{-1}{2}\rangle+e^{i\xi/2} (e^{-i\varphi_1}\sin\frac{\theta_1}{2}+e^{-i\varphi_2}\sin\frac{\theta_2}{2})|\frac{1}{2},\frac{1}{2}\rangle].
\end{equation}
Now, we are interested in  determining the minimum error that could be achieved by measuring the parameter $\xi$.
The  ultimate limit of parameter estimation is determined via unbiased quantum Cram{\'e}r--Rao bound (CRB)  defined as \cite{Giovannetti,Cappellaro} 
\begin{equation}\label{CRB}
\Delta \xi_{CRB}= 1/{\sqrt{F_Q(\rho(\xi))}},
\end{equation}
where $F_Q(\rho(\xi))=Tr[\rho(\xi) L_{\xi}^2]$ is the quantum Fisher information \cite{Giovannetti,Cappellaro}. The symmetric logarithmic derivative $L_{\xi}$ is defined by $\partial_{\xi} \rho(\xi)=(1/2)[\rho(\xi) L_{\xi}+L_{\xi}\rho(\xi) ]$, and $\rho(\xi)$ is the density matrix.

For the  spin-$1/2$ cat state in Eq. (\ref{cat-1/2}), the error of phase measurement $\Delta \xi$ is limited from below by the CRB, for which we find 

\begin{equation}\label{error}
\Delta \xi_{CRB}= \sqrt{\dfrac{2[1+\cos (\frac{\theta_1}{2})\cos (\frac{\theta_2}{2})+\cos (\varphi_1-\varphi_2)\sin (\frac{\theta_1}{2})\sin (\frac{\theta_2}{2})]^2}{[\cos (\frac{\theta_1}{2})+\cos (\frac{\theta_2}{2})]^2[2-\cos (\theta_1)-\cos (\theta_2)+4 \cos (\varphi_1-\varphi_2) \sin (\frac{\theta_1}{2})\sin (\frac{\theta_2}{2})]}}.
\end{equation}
This expression fully captures the CRB of the generic form of the spin-$1/2$ cat state.
Noting that in the Fock basis $|\frac{1}{2},\frac{-1}{2}\rangle=|0,1\rangle$, and $|\frac{1}{2},\frac{1}{2}\rangle=|1,0\rangle$, the state in  Eq.(\ref{cat1/2}) is in form of the single-photon superposition.  Thus, we expect that its performance for phase estimation should be limited by the performance of the single photon N00N state, i.e., $\Delta \xi_{CRB}=1$. As is clear from this  result, the CRB only depends on the phase difference through $\cos (\varphi_1-\varphi_2)$, and it does not depend on each of the phases individually. 
To understand the characteristics of the CRB we consider different situations. First let us take cat states of the form $
\mathcal{N}(|\theta_1,\varphi_1, 1/2\rangle+|\pi-\theta_1,\varphi_2, 1/2\rangle)$. In this case, the CRB reduces to 
\begin{equation}\label{error}
\Delta \xi_{CRB}= \frac{1}{2}\sqrt{\frac{[2+\sin (\theta_1 ) (1+\cos (\varphi_1-\varphi_2))]^2}{[1+\sin (\theta_1 )] [1+\sin (\theta_1 ) \cos (\varphi_1-\varphi_2)]}}.
\end{equation}
From this expression, for $\phi=\varphi_1-\varphi_2=0$ we immediately find $\Delta \xi_{CRB}=1$.   To understand the reason behind this fact we note that the state (\ref{cat1/2}) for $\phi=0$ reduces to
\begin{equation}
|\text{Cat}, 1/2\rangle=\mathcal{N}[(\cos\frac{\theta_1}{2}+\sin\frac{\theta_1}{2} )|\frac{1}{2},\frac{-1}{2}\rangle+(e^{-i\varphi_1}\sin\frac{\theta_1}{2}+e^{-i\varphi_2}\cos\frac{\theta_1}{2})|\frac{1}{2},\frac{1}{2}\rangle].
\end{equation}

\begin{figure}
\centering
\includegraphics[width=4.5 in]{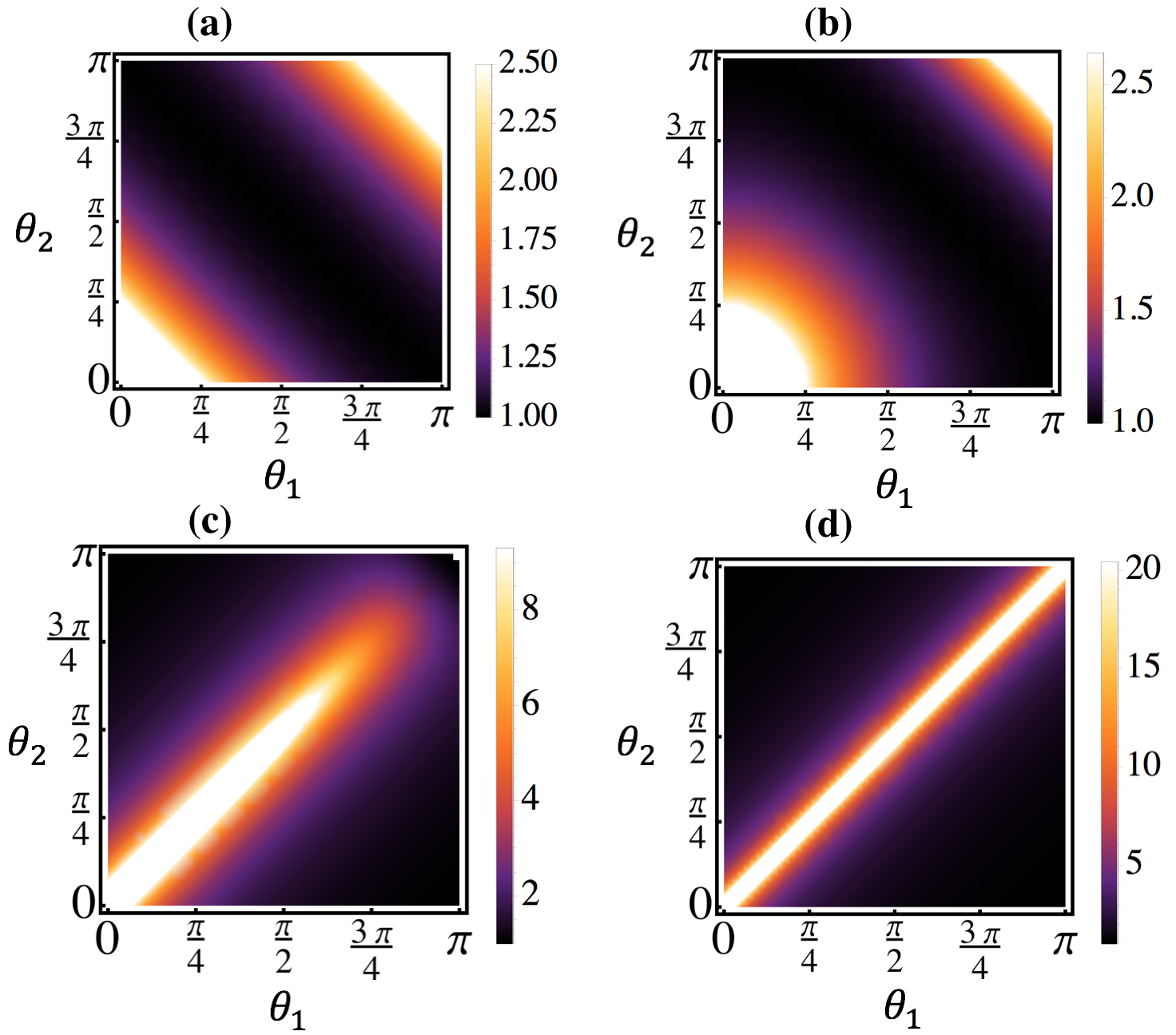}
\caption{
Density of the CRB in terms of $\theta_1$ and $\theta_2$. (a) $\phi=\varphi_1-\varphi_2=0$. (b) $\phi=\pi/2$. (c) $\phi=39 \pi/40 $. (d) $\phi=\pi$.
}
\label{fig:2}
\end{figure}

This state reduces to the single photon N00N state when $\phi=\varphi_1-\varphi_2=0$. This is manifest from Fig. \ref{fig:2}(a). In this plot, the CRB density is given in terms of $\theta_1$ and $\theta_2$ when $\varphi=0$. According to  Fig. (\ref{fig:2}) (a), when $\varphi=0$, the CRB is minimum for $\theta_2=\pi-\theta_1$. At the regions far from $\theta_2=\pi-\theta_1$, the CRB becomes larger. This becomes clear by noting that for  $\varphi=0$, the CRB is simply given by $\Delta \xi_{CRB}=1/|\sin({(\theta_1+\theta_2)}/{2})|$. For instance taking $\theta_1=0$ and $\theta_2=\pi/3$, we get $\Delta \xi_{CRB}= 2$, and for $\theta_2=\pi/2$ one has $\Delta \xi_{CRB}= \sqrt{2}$. In Fig. (\ref{fig:2}) (b) we plot the same density of CRB when $\varphi=\pi/2$. For this scenario, the CRB simplifies as 
\begin{equation}\label{error}
\Delta \xi_{CRB}= \sqrt{\dfrac{2[1+\cos (\frac{\theta_1}{2})\cos (\frac{\theta_2}{2})]^2}{[\cos (\frac{\theta_1}{2})+\cos (\frac{\theta_2}{2})]^2[2-\cos (\theta_1)-\cos (\theta_2)]}}.
\end{equation}
In this case, the CRB is minimum when $\theta_1=0$ and $\theta_2=\pi$, or $\theta_1=\pi$ and $\theta_2=0$. Taking $\theta_1=\theta_2=3\pi/4$, we get $\Delta \xi_{CRB}= 1+\sin ^2\left(\frac{\pi }{8}\right)\approx 1.14645$. Also taking $\theta_1=\theta_2=\pi/2$, we get $\Delta \xi_{CRB}= 3/2\sqrt{2}\approx1.06$. In Fig. (\ref{fig:2}) (c) we plot the same density of CRB when $\varphi=39\pi/40$. The CRB is  still minimum when $\theta_1=0$ and $\theta_2=\pi$, or $\theta_1=\pi$ and $\theta_2=0$; however, it becomes large for the regions far from this condition. Specially, for the region where $\theta_1=\theta_2\lessapprox 3\pi/4$. For instance, taking $\theta_1=\theta_2=\pi/2$, we find $\Delta \xi_{CRB}\approx 12.755$, which is very large compared to the previous subplots. Finally, we take $\varphi=\pi$ in Fig. (\ref{fig:2}) (d) which demonstrates a singularity when $\theta_1=\theta_2$. This could be understood by noting that CRB reduces to $\Delta \xi_{CRB}=1/|\sin({(\theta_1-\theta_2)}/{2})|$ in this scenario. As a result, in all of these plots the when $\theta_1=0$ and $\theta_2=\pi$, or alternatively $\theta_1=\pi$ and $\theta_2=0$, the phase estimation becomes optimal. This is due to the fact that at these cases the cat state become superposition of the north and south poles of the Bloch sphere, and $\varphi_1$ and $\varphi_2$ loose their significance at the poles, making the phase sensitivity independent of these phases and, thus, optimal. 

Now, let us consider the situation where $\theta_1=\theta_2=\pi/2$. Geometrically, this corresponds to the superposition of two coherent states living on the great circle of the Bloch sphere in the plane of $x$ and $y$ axes.
In this case, the CRB simplifies to

\begin{equation}\label{error}
\Delta \xi_{CRB}= \frac{3+\cos (\varphi_1-\varphi_2)}{4|\cos [(\varphi_1-\varphi_2)/2]|}.
\end{equation}
Thus, taking $\varphi_1-\varphi_2=0$, this reduces to $\Delta \xi_{CRB}=1$, and for  $\varphi_1-\varphi_2=\pi$,  $\Delta \xi_{CRB}$ diverges, which is fully in agreement with the observations in Fig. (\ref{fig:2}).

Even though the phase shift on the spin system is usually induced through $e^{i\xi J_z}$, it is shown that phase estimation through $e^{i\xi J_x}$ or $e^{i\xi J_y}$ can also provide helpful estimation information \cite{Sanders}. These phases physically have different meanings.  For instance, considering the Schwinger representation of the spin operators, $e^{i\xi J_z}$ could be understood as a phase shift in an interferometer.  While $e^{i\xi J_x}$ is the beam splitting operator, where its transmittance is determined through $\xi$.

To consider the performance of the state in this platform, we next investigate the phase sensitivity introduced through $e^{i\xi \sigma_x}$.
Following a similar method as the previous case, the CRB can be obtained for  as

\begin{equation}\label{CRB-x1/2}
\Delta \xi_{CRB}= \left[1-\frac{[\cos (\frac{\theta_1}{2})+\cos (\frac{\theta_2}{2})]^2[\cos (\varphi_1)\sin (\frac{\theta_1}{2})+\cos (\varphi_2)\sin (\frac{\theta_2}{2})]^2}{[1+\cos (\frac{\theta_1}{2})\cos (\frac{\theta_2}{2})+\cos (\varphi_1-\varphi_2)\sin (\frac{\theta_1}{2})\sin (\frac{\theta_2}{2})]^2}\right]^{-1/2}.
\end{equation}
From this result, we immediately conclude that the necessary and sufficient condition for HL phase sensitivity, here, is $\cos (\varphi_1)\sin (\frac{\theta_1}{2})+\cos (\varphi_2)\sin (\frac{\theta_2}{2})=0$.  It  also becomes clear that for $\theta_1=\theta_2=0$, (or alternatively $\theta_1=\theta_2=\pi$) the sensitivity of the cat state becomes optimal. We note that, at these extreme points, the state reduces to the south or north pole of the Bloch sphere, with the resultant states $|\frac{1}{2},-\frac{1}{2}\rangle$ or  $|\frac{1}{2},\frac{1}{2}\rangle$.  It is insightful to note that these states correspond to $|0\rangle |1\rangle$ and $|1\rangle |0\rangle$ in the Fock state representation. Even though these states provide no information about the phase accuracy for $e^{i\xi \sigma_z}$, they provide the best accuracy about $\xi$ for $e^{i\xi \sigma_x}$. This is due to the fact that sending a single photon Fock state through a beam splitter transforms it into a superposition state. 

We note that, unlike the previous scenario, in this case, the CRB is controlled by each of the phases individually rather than only the phase difference being important. An interesting choice would be $\{\varphi_1,\varphi_2\}=\{\pi/2,3 \pi/2\}$, which also results in the optimal phase estimation. This becomes more insightful if we note that taking $\varphi=\pi/2 (3 \pi/2)$ demonstrates coherent states on the great circle in  $y$ and $z$  plane of the Bloch sphere. Therefore, if we choose the superposition of any two coherent states on this circle, we attain HL sensitivity for $\xi$.

\begin{figure}[h]
\centering
\includegraphics[width=4.5 in]{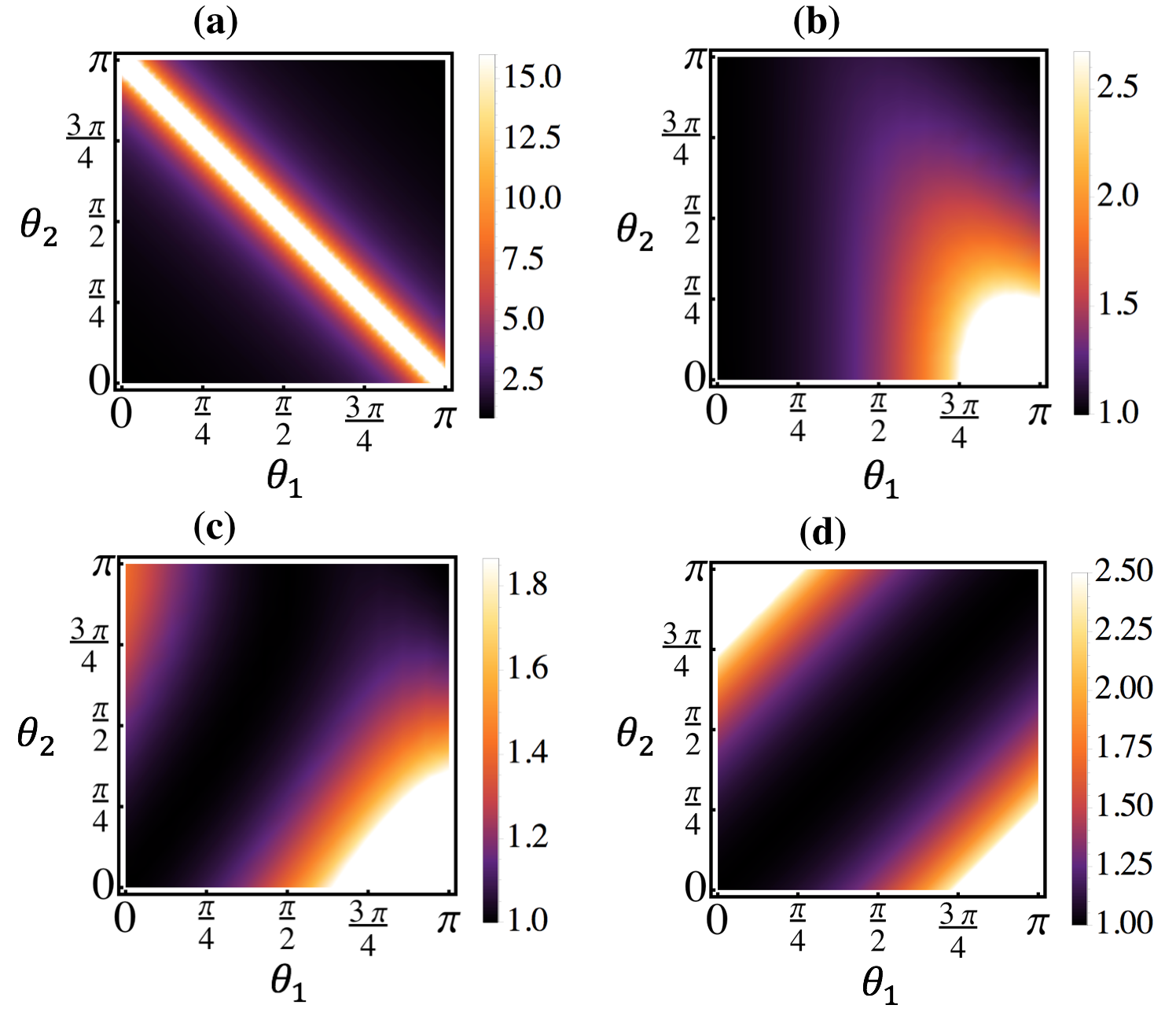}
\caption{
Density of the CRB in terms of $\theta_1$ and $\theta_2$. We set  $\varphi_1=0$ in these plots. (a) $\varphi_2=0$. (b) $\varphi_2=\pi/2$. (c) $\varphi_2=3 \pi/4$. (d) $\varphi_2=\pi$.
}
\label{fig:3}
\end{figure}

Fig. (\ref{fig:3})  presents CRB in terms of $\theta_1$ and $\theta_2$ for various $\varphi_2$. In all of the sub-figures we have chosen $\varphi_1=0$. In Fig. \ref{fig:3}(a) we have considered $\varphi_2=0$, where the CRB becomes optimal for  $\theta_1=\theta_2=0,\pi$, and it is maximal for $\theta_1+\theta_2=\pi$. To understand this we note that setting $\varphi_1=\varphi_2=0$ in Eq. (\ref{CRB-x1/2}), the CRB reduces to $\Delta \xi_{CRB}=1/{|{\cos[(\theta_1+\theta_2)/2]}|}$. The same  plot with $\varphi_2=\pi/2$, is given in Fig. \ref{fig:3}(b). Accordingly, $\theta_1=0$, or $\theta_1=\theta_2=\pi$, provides optimal CRB. To have a better insight into this figure we should note that for this situation the CRB reduces to
\begin{equation}\label{}
\Delta \xi_{CRB}= \left[1-\frac{[\cos (\frac{\theta_1}{2})+\cos (\frac{\theta_2}{2})]^2[\sin (\frac{\theta_1}{2})]^2}{[1+\cos (\frac{\theta_1}{2})\cos (\frac{\theta_2}{2})]^2}\right]^{-1/2}.
\end{equation}
Thus, taking $\theta_1=0$, or $\theta_1=\theta_2=\pi$, in this equation, gives $\Delta \xi_{CRB}=1$. Assuming $\theta_1=\theta_2=\theta$, we can further simply this equation to 

$$
\Delta \xi_{CRB}=\frac{\sqrt{2}[3+\cos ({\theta})]}{\sqrt{15+12\cos ({\theta})+5\cos ({2\theta})}},
$$
 resulting in $\Delta \xi_{CRB}=\frac{3}{\sqrt{5}}$ for $\theta=\pi/2$. 
In Fig. \ref{fig:3}(c), we take $\varphi_2=3\pi/4$. Accordingly, the phase estimation can reach optimal value for specific  choices of the parameters, and it diverges for  $\theta_1=\pi$. If we choose $\theta_2=0$  the CRB shall simplify to $\Delta \xi_{CRB}=1/{{\cos(\theta_1/2)}}$. Thus, as the plot presents the CRB increases by increasing $\theta_1$. Alternatively, taking $\theta_1=0$, the CRB simplifies to $\Delta \xi_{CRB}=2/\sqrt{3+\cos(\theta_2)}$, ranging between 1 and $\sqrt{2}$.

Now, we take $\varphi_2=\pi$, in 
Fig. \ref{fig:3}(d), which presents optimal phase estimation for $\theta_1=\theta_2$. This could be understood by noting that $\Delta \xi_{CRB}=1/{|{\cos[(\theta_1-\theta_2)/2]}|}$, for this situation. From Fig. (\ref{fig:3}) it is clear that for $\theta_1=\theta_2=0,\pi$ the phase sensitivity becomes independent of $\varphi_1$ and $\varphi_2$, and it gives $\Delta \xi_{CRB}=1$. As the last case, we take $\theta_1=\theta_2=\pi/2$. The CRB reduces then to
\begin{equation}\label{CRB-x1/2}
\Delta \xi_{CRB}= \left[1-\frac{4[\cos (\varphi_1)+\cos (\varphi_2)]^2}{[3+\cos (\varphi_1-\varphi_2)]^2}\right]^{-1/2}.
\end{equation}

Thus, $\varphi_1=\varphi_2=0$ gives the maximal, and $\varphi_1=\varphi_2=\pi$ gives the minimal value for the CRB, as can be seen from the plots.





\section{Quantum Metrology with Spin-1  Cat States}

Now, we generalize the investigations of the previous section to the cat states with $j=1$. To this end,  
taking $j=1$, the coherent state reduces to

\begin{equation}\label{SCS1}
|\theta,\varphi, 1\rangle
=\cos^2(\theta/2)
|1,-1\rangle+\frac{1}{\sqrt{2}}e^{i\varphi}\sin({\theta})|1,0\rangle+e^{2i \varphi} \sin^2(\theta/2)|1,1\rangle,
\end{equation}

Noting that in the Fock basis $|1,-1\rangle=|0,2\rangle$,  $|1,0\rangle=|1,1\rangle$, and $|1,1\rangle=|2,0\rangle$, the state in  Eq.(\ref{SCS1}) is in form of the two photon superposition. Similar to the previous section, we consider a general form of the superposition of two spin-1 coherent states as 
\begin{equation}
|\text{Cat}, 1\rangle=\mathcal{N}(|\theta_1,\varphi_1, 1\rangle+|\theta_2,\varphi_2, 1\rangle).
\end{equation}
Here, $\mathcal{N}$ is the normalization factor of the spin cat states.

\begin{figure}[h]
\centering
\includegraphics[width=4.5 in]{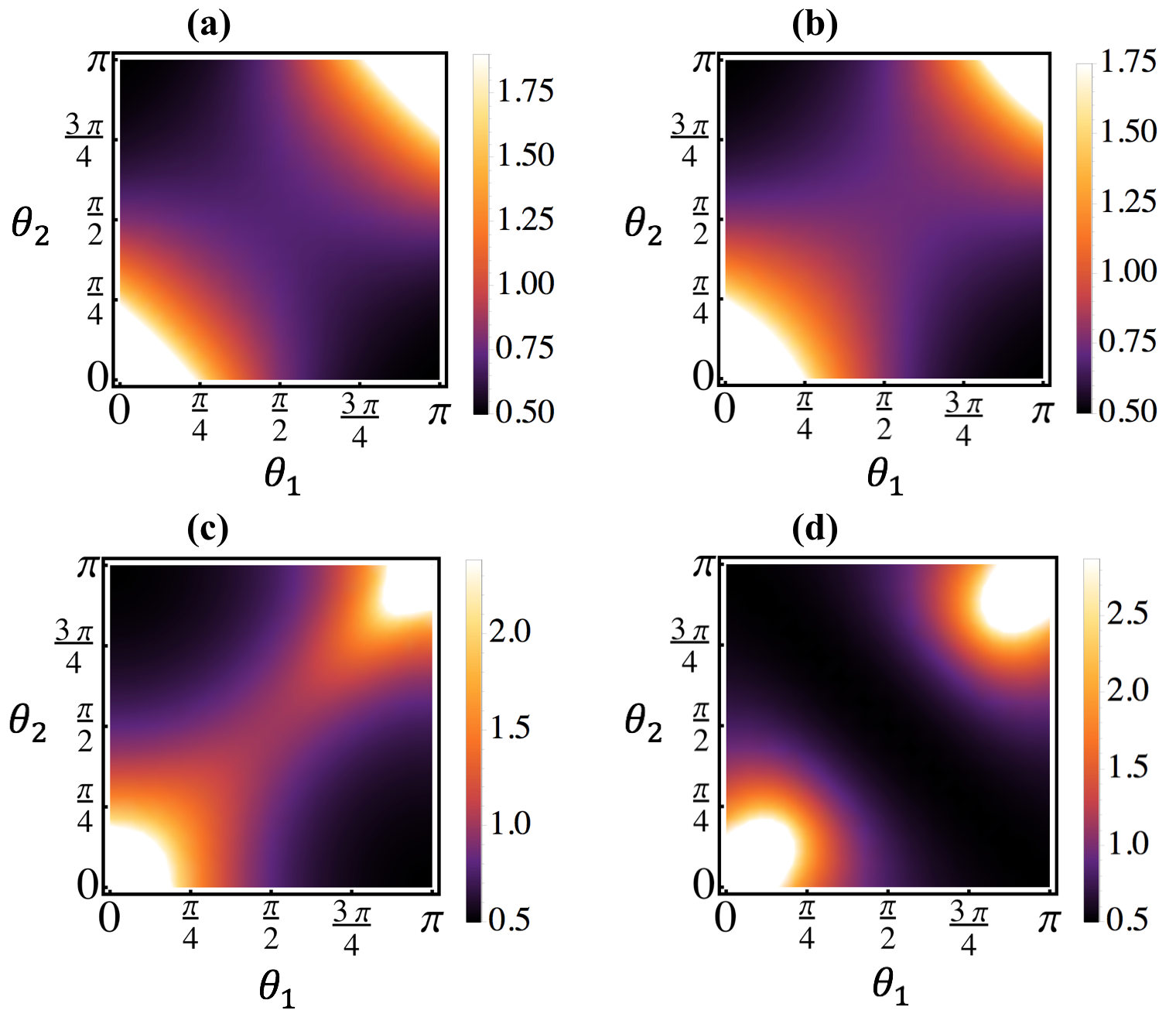}
\caption{
Density of the CRB in terms of $\theta_1$ and $\theta_2$. (a) $\phi=0$. (b) $\phi=\pi/2$. (c) $\phi=3 \pi/4$. (d) $\phi=\pi$.
}
\label{fig:4}
\end{figure}

We analyze the performance of the state $|\text{Cat}, 1\rangle$ for quantum metrology when the system has accumulated a phase through $e^{i\xi J_z}$.  For such a phase shift, we are going to provide analytic results for various scenarios. It turns out that the CRB depends on the $\cos(\varphi_1-\varphi_2)$, and not each of the phases individually. This is similar to the spin-1/2 case, and hence we can conclude that the metrological performance of the cat states through $e^{i\xi J_z}$ is controlled by the phase differences. We present the CRB density 
as a function of $\theta_1$ and $\theta_2$ for various $\phi=\varphi_1-\varphi_2$  in Fig. (\ref{fig:4}). Accordingly, the phase difference has a significant role in the performance of the states.
We take $\phi=0$, in Fig. \ref{fig:4}(a). In this case, the cat state reduces to 
\begin{equation}
|\text{Cat}, 1\rangle=\mathcal{N}(|\theta_1,\varphi, 1\rangle+|\theta_2,\varphi, 1\rangle).
\end{equation}
For such a superposition state,  the CRB of the system could be found to be

\begin{equation}\label{CRB-1a}
\Delta \xi_{CRB}= \left[\frac{2 (\cos (\theta_1-\theta_2)+3)^2
}{\mathcal{A}-8\mathcal{B}+ 2\mathcal{C}-18 \mathcal{D}+30}\right]^{1/2},
\end{equation}
where,
\begin{align}
\mathcal{A}&=\cos (3 \theta_1-\theta_2)+\cos (3 \theta_2-\theta_1),
\\
\mathcal{B}&=\cos (2 \theta_1)+ \cos (2 \theta_2),
\\
\mathcal{C}&=\cos (2 (\theta_1-\theta_2)),
\\
\mathcal{D}&=\cos (\theta_1+\theta_2).
\end{align}
As was expected, $\Delta \xi_{CRB}$ is symmetric in terms of  $\theta_1$ and $\theta_2$. Particularly, if one of them is chosen to be 0 and the other is set to be $\pi$, we attain HL sensitivity $\Delta \xi_{CRB}=1/2$. This observation is expected due to the fact that one of the coherent states, at these cases, is at the north and the other at the south pole of the  Bloch sphere. However, superposition of the poles provide N00N states, providing HL precision. This is also clear from Fig. \ref{fig:4}(a). According to this figure, $\Delta \xi_{CRB}$ is optimal only at these specific choices. If we consider $\theta_1=\theta_2$, then $\Delta \xi_{CRB}=1/\sqrt{2}|\sin(\theta_1)|$. At this scenario, the best performance is achieved for $\theta_1=\pi/2$, and  $\Delta \xi_{CRB}$ diverges for   $\theta_1=0$ or $\pi$. This is clearly shown in Fig. \ref{fig:4}(a). The other interesting situation in this class of cat states is considering $\theta_2=\pi-\theta_1$. From Fig. \ref{fig:4}(a), it can be realized that the $\Delta \xi_{CRB}$ provides a better sensitivity along this line.  With this assumption, the explicit form of the cat state will be
\begin{equation}
|\text{Cat}, 1\rangle=\mathcal{N}(|\theta_1,\varphi, 1\rangle+|\pi-\theta_1,\varphi, 1\rangle).
\end{equation}
This type of state is considered in \cite{Huang1,Huang2}. The CRB provided by this state is given in Fig. \ref{fig:4}(a).
For such a superposition state, we find for the CRB of the system
\begin{equation}
\Delta \xi_{CRB}= \left[\frac{3- \cos (2\theta_1)
}{8}\right]^{1/2}.
\end{equation}

Accordingly, for $\theta_1=0,\pi$ we attain HL sensitivity, while the minimum sensitivity happens for $\theta_1=\pi/2$ which gives $\Delta \xi_{CRB}=1/\sqrt{2}$, which is equal to the shot-noise-limit. Thus, $\Delta \xi_{CRB}$ is always better than shot-noise-limit, except for $\theta_1=\pi/2$.

In Fig. \ref{fig:4}(b) we consider the performance of the cat states of the form
\begin{equation}
|\text{Cat}, 1\rangle=\mathcal{N}(|\theta_1,\varphi, 1\rangle+|\theta_2,\varphi+\pi/2, 1\rangle).
\end{equation}
Geometrically, the two coherent states are on the two orthogonal circles on the Bloch sphere. For instance, if one of them is on the circle in the plane of $x$ and $z$ axes, the other must be on the circle crossing $y$ and $z$ axes.
 For this cat state, $\Delta \xi_{CRB}$ could be found to be 
 \begin{equation}\label{CRB-1b}
\Delta \xi_{CRB}= \left[\frac{ \mathcal{A}^2
}{\mathcal{A}\mathcal{B}-\mathcal{C}}\right]^{1/2},
\end{equation}
where,
\begin{align}
\mathcal{A}&=\cos (\theta_1)+\cos (\theta_2)+2,
\\
\mathcal{B}&=\cos (2 \theta_1)+\cos (2 \theta_2)-8 \cos (\theta_1) \cos (\theta_2)+6,
\\
\mathcal{C}&=4 (\cos (\theta_1) \cos (\theta_2)-1)^2.
\end{align}
 It is easy to verify that  for $\theta_1=\theta_2=0,\pi$ $\Delta \xi_{CRB}$ diverges since $\mathcal{A}\mathcal{B}-\mathcal{C}=0$. While, if one of the angles is 0 and the other is $\pi$, we immediately attain the HL sensitivity. These could also be seen from Fig. \ref{fig:4}(b). Now, let us consider the specific case of $\theta_1=\theta_2$. This geometrically indicates the situation where both coherent states are on the same circle that is orthogonal to $z$ axe.  For this state, we find $\Delta \xi_{CRB}=1/|\sin (\theta_1)|$. Clearly, this is above the shot-noise limit, and hence, the state does not provide any quantum enhancement for metrological applications.

To provide another situation, we considering $\theta_2=\pi-\theta_1$. From Fig. \ref{fig:4}(b), it can be realized that  $\Delta \xi_{CRB}$ provides a better sensitivity along this line. Here, the explicit form of the cat state will be
\begin{equation}
|\text{Cat}, 1\rangle=\mathcal{N}(|\theta_1,\varphi, 1\rangle+|\pi-\theta_1,\varphi+\pi/2, 1\rangle).
\end{equation}
 For this cat state, $\Delta \xi_{CRB}$ could be found as
 \begin{equation}\label{CRB-1b}
\Delta \xi_{CRB}= \left[\frac{8}{12 \cos (2 \theta_1)-\cos (4 \theta_1)+21}\right]^{1/2}.
\end{equation}
This reduces to HL metrology for $\theta_1=0,\pi$, while it gives $\Delta \xi_{CRB}=1$ for $\theta_1=\pi/2$. 

In Fig. \ref{fig:4}(c), we consider the performance of a cat state given by

\begin{equation}
|\text{Cat}, 1\rangle=\mathcal{N}(|\theta_1,\varphi, 1\rangle+|\theta_2,\varphi+3\pi/4, 1\rangle).
\end{equation}
Similar to the previous cases, the sensitivity is larger as the difference between $\theta_1$ and $\theta_2$ is bigger. This means that when one of the coherent states is close to the north and the other to the south pole of the Bloch sphere, parameter estimation accuracy is higher. For instance, let us consider   $\theta_1 = \pi/4$ and $\theta_2 = 3\pi/4$. Then we get $\Delta \xi_{CRB} \approx 0.525$, which is far below the shot-noise limit. Of course, further moving the coherent states towards the poles will improve the performance.

Finally, in Fig. \ref{fig:4}(d), we plot $\Delta \xi_{CRB}$ for the cat state
\begin{equation}
|\text{Cat}, 1\rangle=\mathcal{N}(|\theta_1,\varphi, 1\rangle+|\theta_2,\varphi+\pi, 1\rangle).
\end{equation}
To understand the performance we obtain  $\Delta \xi_{CRB}$ for this cat state to be
\begin{equation}\label{CRB-1a}
\Delta \xi_{CRB}= \left[\frac{2 (\cos (\theta_1+\theta_2)+3)^2
}{\mathcal{A}-8\mathcal{B}+ 2\mathcal{C}-18 \mathcal{D}+30}\right]^{1/2},
\end{equation}
where,
\begin{align}
\mathcal{A}&=\cos (3 \theta_1+\theta_2)+\cos (\theta_1-3 \theta_2),
\\
\mathcal{B}&=\cos (2 \theta_1)+ \cos (2 \theta_2),
\\
\mathcal{C}&=\cos (2 (\theta_1+\theta_2)),
\\
\mathcal{D}&=\cos (\theta_1-\theta_2).
\end{align}

To have a better insight into the behavior of  $\Delta \xi_{CRB}$ along the line $\theta_1 =\theta_2$, in Fig. \ref{fig:4}(d), we note that with this constrain $\Delta \xi_{CRB}$ reduces to 

\begin{equation}
\Delta \xi_{CRB}= \left[\frac{3+\cos (2\theta_1)
}{4 |\sin(\theta_1)|}\right].
\end{equation}
Therefore,  $\Delta \xi_{CRB}$ diverges for $\theta_1 =0,\pi$, and it reduces to HL for $\theta_1 =\pi/2$. The other interesting scenario would be  to consider $\Delta \xi_{CRB}$ along the line $\theta_1 =\pi-\theta_2$. For this case, the cat state reduces to

\begin{equation}
|\text{Cat}, 1\rangle=\mathcal{N}(|\theta_1,\varphi, 1\rangle+|\pi-\theta_1,\varphi+\pi, 1\rangle).
\end{equation}
 We note that this is a superposition of two antipodal points on the Bloch sphere, which we refer to it as the antipodal cat state. In this case, the two coherent states become orthogonal. Interestingly enough, $\Delta \xi_{CRB}$ provides HL sensitivity for any choices of the parameters in this case. Therefore,  antipodal cat state provides HL metrology for $j=1$.

As a result, in all of these plots, when $\theta_1=0$ and $\theta_2=\pi$, or alternatively $\theta_1=\pi$ and $\theta_2=0$, the phase estimation becomes optimal. This is due to the fact that in these cases, the cat state becomes a superposition of the north and south poles of the Bloch sphere, and $\varphi_1$ and $\varphi_2$ lose their significance at the poles, making the phase sensitivity independent of these phases and, thus, optimal.





\section{Conclusion}

To summarize, we have considered the quantum metrological performance of spin coherent states superposition and identified conditions for measurements with the  Heisenberg-limit precision. We demonstrated that the operator's choice, which introduces the parameter for the estimation, plays a significant role in the performance of the quantum probe. We derive closed-form analytical descriptions for the performance of spin cat states. 
These findings suggest that careful control of parameters is necessary for achieving Heisenberg-limit precision and provide insightful information on the geometry of the specific coherent state superposition on the Bloch sphere and its relevance to the performance of the states for parameter estimations.

\end{document}